\documentclass{article}

\usepackage[square,sort,comma,numbers]{natbib}
\usepackage[final]{nips_2016}
\usepackage[utf8]{inputenc} 
\usepackage[T1]{fontenc}    
\usepackage{url}            
\usepackage{booktabs}       
\usepackage{amsfonts}       
\usepackage{nicefrac}       
\usepackage{microtype}      
\usepackage{soul}
\usepackage{color}
\usepackage{soul}
\title{MusculoSkeletal Modeling Using Kinect Data For Telerehabilitation}

\author{Rajat Kumar Das, Soumya Ranjan Tripathy, Kingshuk Chakravarty, \\
\textbf{Debatri Chatterjee, Aniruddha Sinha, Rupam Chaudhury} \\
TCS Research, Tata Consultancy Services Ltd.,
Kolkata, India \\
\texttt{das.rajat@tcs.com, sr.tripathy@tcs.com, kingshuk.chakravarty@tcs.com,} \\ \texttt{debatri.chatterjee@tcs.com, aniruddha.s@tcs.com, r.chaudhury@tcs.com}
}

\begin{document}

\maketitle

\begin{abstract}
Balance, gait and postural control are some of the key factors in determining the overall stability of an individual. Several highend and costly solutions exist to perform movement analysis in clinical settings. OpenSim is a tool which uses 39 marker positions, obtained from such highend solutions like VICON or equivalent multicamera setup, for the analysis of inverse kinematics and inverse dynamics. However, an affordable solution for deriving musculoskeletal  joint kinematics parameters using a low cost Kinect device is of immense importance. In this paper, we initially study the feasibility of using OpenSim tool on 20 joint locations of human being, obtained from Kinect data. Next, we analyze the various joint forces and torques experienced during a Single Limb Stance (SLS) exercise performed by healthy subjects in normal, overweight and obese categories. Results indicate that a subset of parameters related to forces and torque in hip, lumber and pelvis are the most important ones that contribute significantly in maintaining static balance in SLS. Statistical analysis demonstrates that the pelvis list and tilt moments are the key biomarkers for maintaining the statibility in SLS, thus leading to a possibility of personalizing the therapy in tele-rehabilitation. 
\end{abstract}


\section{Introduction} Humans are bipeds and their locomotion over the ground using the lower limbs results in different postures like standing, walking, running etc. 
The ability to maintain a natural upright posture and control its stability is dependent on both the nervous system and locomotor system (muscles, bones, joints). However due to aging and various pathological conditions including stroke, the degeneration of these control system happens which intern increases the risk of gait disorder, injury, fall and immobility of the normal life. Stroke survivors have serious falls within a year of their stroke. Balance disorders are most common health problem of patients over age 70. Therefore there is a substantial need for screening and to have some preventive biomarkers for quantifying human balance, gait kinematics estimation and joint force estimation of the lower body. This system will also help to design personalized rehabilitation or therapies too.

Therefore our motivation is to capture human gait through a low cost motion sensing device like Kinect and do inverse kinematics and inverse dynamics, so that from posture trajectories we can derive the joint forces and torques. 
Then an intelligent machine learning system will help us understand the consequences between the joint forces for different sets of people like normal, overweight and obese groups. It will also help us to find bio-markers responsible for the stability control. We have used the SLS as one of the static balance exercise to analyze the joint forces and torques \cite{chakravarty2016quantification}. Further these biomarkers can be used by clinicians for rehabilitation through personalized therapy in order to improve the stability and reduce fall risk.


\section{Prior art} The human body posture can be described using a simple variant of an inverse pendulum model, by assuming ankle as hinge joint and hip as the pendulum. This system can be described by the pendulums equation of motion which can be solved either by free body diagram (FBD), Newton’s-Eulers or Lagrangian like approaches \cite{kot2014modeling}, \cite{winter1998stiffness},\cite{winter1995human},\cite{stimac1999standup}. This single pendulum approach can be further improved by making it a series of inverse pendulum \cite{pinter2008dynamics}. However, these approaches assume the whole body as a point mass. Further improvement happens with Reaction Mass Pendulum (RMP) model, where it assumes extended rigid body mass \cite{orin2013centroidal},\cite{dutta2010human}. However todays musculoskeletal modeling software enables us to use a human body like structure where the human body properties like muscle, joints and different body elements, tendons are defined by expert biologists and the equations for the underlying dynamics are solved by Newton's-Eulers approaches \cite{delp2007opensim}. These models are based on more realistic assumptions compare to inverse pendulum model for deriving inverse kinematics and inverse dynamics related solution. On using these modeling softwares the whole process becomes less error prone and takes less execution time. There are various licensed softwares available like SIMM \cite{delp2000computationalSIMM}, LifeMod  \footnotemark[1], Anybody \footnotemark[2] developed by Stanford University, LifeModeler and Anybody Technology respectively. In this paper, we have used Opensim \cite{delp2007opensim} which is freeware developed by Stanford University. A recent comparison is done on Kinect and VICON based human motion analysis \cite{jun2003comparativeKinect}, however it does not focus on the findings related to static balance analysis. Our focus is on analyzing normal, overweight and obese subjects using Kinect sensor and Wii balance board data along with OpenSim for finding biomarkers related to SLS balance.

\footnotetext[1]{\url{http://www.lifemodeler.com/LM_Manual/modeling_motion.shtml}}
\footnotetext[2] {\url{http://www.anybodytech.com/652.0.html}}

\section{Methodology} 
In order to attain the goal, we have chosen Kinect Xbox 360 from Microsoft, as a motion sensing device, due to its affordability and portability compared to highend systems like VICON, which is very costly and also require skilled person to set up the marker positions \cite{sinha2016accurate}. The 3D co-ordinates (x,y,z) of various joints of a human is obtained as a time series for different postures. We have used OpenSim musculoskeletal modeling software \cite{delp2007opensim}, which is capable of doing inverse kinematics and inverse dynamics to obtain different joints forces and moments. From available OpenSim model list, we have used \textit{gait2392} Simbody model. This primarily contains the lower extremity model with two legs and a lumped torso segment. It includes 23 degrees of freedom and 92 muscle-tendon actuators and 39 markers. In order to feed the Kinect data (20 joint coordinates as a time series) into the model, we need to follow few steps like (i) marker set reduction, (ii) position adjustment with respect to Kinect, (iii) coordinate transformation from Kinect coordinate to OpenSim coordinate system and (iv) scaling. The scaling is performed based on a measured distances between x-y-z marker locations. The marker locations are usually obtained using motion capture equipment. An unscaled model has a set of virtual markers placed in the same anatomical locations as the experimental markers. The dimensions of each segment in the model are scaled so that the distances between the virtual markers match the distances between the experimental markers. Once the scaled version is obtained we get a personalised musculoskeletal model from a generic one. 

Then the next step is to do the inverse kinematics which find the joint angles and positions that best match the experimental kinematics recorded during SLS. The Opensim inverse kinematics tool minimizes the least square error between the computed generalized coordinated with the available experimental marker coordinates. Then the generalized coordinate values are fed into the inverse dynamics toolkit along with external force values (center of pressure and ground reaction force) to calculate the generalized internal force values that causes a particular motion in the human body, in our case for SLS. These generalized force values are then statistically analyzed to derive the reduced set of parameters (forces and moments) responsible for significantly categorizing different sets of people like normal, overweight, obese.
 
\section{Experimental Procedure} 
Every participants in the laboratory are explained with the anonymized data collection protocol along with an initial demonstration. The protocol is cleared by our Institute Review Board. An informed consent is taken from each participants. Then the weight and height is measured using weight machine and measuring tape. Next each participant stands in front of Kinect and on top of the Wii balance board. The experiment starts with a 10 second double limb standing phase, which is required for scaling. After that one limb is lifted for the SLS phase. The duration of the SLS is left to participant's choice, where they can stand as long as they are comfortable with maintaining the static balance. The data captured during SLS phase is used for the calculation of inverse kinematics and inverse dynamics. During the experiment the Kinect and Wii board data are being synchronously captured by our in house data acquisition system. Here we have collected a total of 30 participants data with 10 belonging to each group from normal, overweight and obese, according to BMI index. The age group lies between 25 to 35 years.

\section{Results and Analysis}
The available standard full body human model of the OpenSim software has 39 markers in different body parts. In order to analyze the Kinect data in OpenSim, with only 20 markers, first we validate the feasibility of using the output of the inverse dynamics as obtained from reduced number of markers with OpenSim. The root mean square error is computed between the parameters obtained with the available OpenSim data by taking all 39 markers and comparing with a reduced set of 20 markers which coincides with the Kinect data. We have found that with the reduced marker set of 20, the average root mean squared error for the force and moment related parameters for all joint is 16.45\%. This error indicates the feasibility of using inverse dynamics of OpenSim with the Kinect data where a reduction of approximately 50\% marker data leads to an error of only 17\%  approximately.

For every participant, first 10 seconds of standing data obtained from Kinect is used for scaling purpose. The dimensions of each segment in the model are scaled so that the distances between the virtual markers match the distances between the experimental markers. After the scaling from a generic model is done, a personalized model is generated. Next step is inverse kinematics (IK), where for each time frame of motion, OpenSim IK tool computes generalized coordinate values, which re-position the model in a pose that best matches the experimental markers and the coordinate values computed for that step. Finally, the Inverse dynamics tool is used to compute generalized forces for each joint, responsible for the given motion. Inverse dynamics require generalized coordinate values and the external force which is ground reaction force in our case. It also requires the center of pressure (COP) and ground reaction force (GRF) as input which we get from the Wii balance board. For all the three groups namely normal, overweight and obese, we run the above steps to get the generalized forces for every joints like hip, pelvis, knee etc. We perform the ANOVA analysis to evaluate the most significant parameter set, representing the generalized forces and moments, responsible to separate the three groups (normal, overweight and obese). Table \ref{sample-table} represents the top seven parameters having a high F value with p < 0.05, indicating the discriminative power of the parameters for the SLS performed with lifting the right leg. It can be seen that the moments associated with the right hip, lumber and pelvis represent as a significant biomarker for distinguising the groups. Among this the pelvis list moment (F=2867) plays the major role in maintaining the static balance in SLS followed by pelvis tilt moment. Medical practitionars and physiotherapists can use this information and methodology to personalize the therapy in order to improve the static balance.

\begin{table}[t]
  \caption{Significant parameters based on ANOVA analysis for normal, overweight and obese}
  \label{sample-table}
  \centering
  \begin{tabular}{lll}
    \toprule
    Inverse dynamics    & p value     & F value \\
    \midrule
    right hip abduction moment & 0  & 316.92     \\
    right hip rotation moment  & 0 & 140.01      \\
    lumber bending moment  & 0       & 340  \\
    lumber rotation moment  & 0       & 202.33  \\
    pelvis rotation moment  & 0       & 512.67  \\
    pelvis tilt moment  & 0       & 578.44  \\
    pelvis list moment  & 0       & 2867  \\
    
    \bottomrule
  \end{tabular}
\end{table}
 
\section{Conclusion and future scope}
Joint forces and moments play a major role in maintaining the balance of an individual. Experiment is performed on healthy subjects with normal, overweight and obese categories to analyze the joint forces and moments during SLS. The human motion data is captured using Kinect followed by musculoskeletal modeling using OpenSim to derive the joint forces and moments. ANOVA analysis is performed to derive the significant parameters (pelvis list and tilt moment) acting as biomarkers for maintaining the balance in SLS. Thus subjects with high fall risk can be targetted with exercises to strengthen muscles to improve these moments. In future, we need to perform clinical trials to validate the same on patient data, using longitudinal study and for other exercises related to balance, gait and range of motion.

\section{Acknowledgments}
The authors are immensely grateful to all the participants for their valuable time and participation in the study.
\bibliographystyle{unsrt}
\bibliography{nips}
%
%
%

%

\end{document}